\documentclass[%
 reprint,
superscriptaddress,
 amsmath,amssymb,
 aps,
prb,
floatfix,
]{revtex4-2}

\usepackage{appendix}
\usepackage{lipsum}
\usepackage{physics}
\usepackage{bbm}
\usepackage{graphicx}
\usepackage{bm}

\usepackage{xcolor}

\usepackage{verbatim}

\begin{document}

\preprint{APS/123-QED}

\title{Optical absorption in two-dimensional materials with tilted Dirac cones}

\author{Andrew Wild}
\email{A.Wild@exeter.ac.uk}
\author{Eros Mariani}
\email{E.Mariani@exeter.ac.uk}
\author{Mikhail E. Portnoi}
\email{M.E.Portnoi@exeter.ac.uk}
\affiliation{%
 Physics and Astronomy, University of Exeter, Stocker Road, Exeter EX4 4QL, United Kingdom
}%

\date{\today}
\begin{abstract}
The interband optical absorption of linearly polarised light by two-dimensional (2D) semimetals hosting tilted and anisotropic Dirac cones in the bandstructure is analysed theoretically. Super-critically tilted (type-II) Dirac cones are characterised by an absorption that is highly dependent on the incident photon polarisation and frequency, and is tunable by changing the Fermi level with a back-gate voltage. Type-II Dirac cones exhibit open Fermi surfaces and large regions of the Brillouin zone where the valence and conduction bands sit either above or below the Fermi level. As a consequence, unlike their sub-critically tilted (type-I) counterparts, type-II Dirac cones have many states that are Pauli blocked even when the Fermi level is tuned to the level crossing point. We analyse the interplay of the tilt parameter with the Fermi velocity anisotropy, demonstrating that the optical response of a Dirac cone cannot be described by its tilt alone. As a special case of our general theory we discuss the proposed 2D type-I semimetal 8-$Pmmn$ Borophene. Guided by our in-depth analytics we develop an optical recipe to fully characterise the tilt and Fermi velocity anisotropy of any 2D tilted Dirac cone solely from its absorption spectrum. We expect our work to encourage Dirac cone engineering as a major route to create gate-tunable thin-film polarisers. 
\end{abstract}
\maketitle

\section{Introduction}

Two-dimensional Dirac semimetals (DSM) host low energy electronic excitations described by Dirac cones with linear dispersions. The most famous example of a DSM is graphene hosting isotropic Dirac cones. However, in general, DSMs can host tilted Dirac cones\,\cite{Bernevig} in the band structure. The Dirac cones in graphene belong to the wider class of Dirac cones with sub-critical tilt (type-I), meaning that the Fermi surface is elliptical and compact. Beyond type-I Dirac cones there exists DSMs with open Fermi surfaces: critically tilted (type-III) Dirac cones with parabolic Fermi surfaces and super-critically tilted (type-II) Dirac cones with hyperbolic Fermi surfaces. There exists a variety of candidate tilted Dirac cones\,\cite{2DDSM} in different systems  such as: partially hydrogenated graphene\,\cite{TiltedCone1}, warped graphene\,\cite{TiltedCone2}, organic conductor $\alpha$-(BEDT-TTF)$_2\text{I}_3$\,\cite{OrganicTiltedCone1}, 8-$Pmmn$ Borophene\,\cite{Borophene1,Borophene2}, planar arrays of carbon nanotubes\,\cite{CNTArray}, artificial graphenes\,\cite{Mann2018,Mann2020} and others\,\cite{TypeIIMonolayer,DiracConeAdditional1,DiracConeAdditional2,DiracConeAdditional3,DiracConeAdditional4,DiracConeAdditional5,DiracConeAdditional6}.

The Dirac cones in the band structure of graphene are responsible for its universal sheet absorbance at normal incidence of $2.3\%$\,\cite{GrapheneAbsTheory1,GrapheneAbsTheory2} independent of polarisation\cite{GrapheneAbsExperiment}. The absorption is frequency independent beyond a cut-off frequency that can be tuned by modulating the Fermi level via a back-gate voltage. Although the absorption spectra of some type-I Dirac materials have been investigated\,\cite{Verma,Herrera,Nishine,Suzumura,Jafari,PhysRevB.103.125425,PhysRevB.103.165415}, a general theory of absorption of generic 2D DSMs with Dirac cones of arbitrary tilt and Fermi velocity anisotropy has not been formulated. Similarly, the optical conductivity of tilted 3D Dirac/Weyl cones\,\cite{Carbotte} qualitatively differs from the 2D case due to the increased dimensionality of the materials. Two-dimensional Dirac semimetals hosting tilted Dirac cones are drawing ever growing experimental and theoretical interest. Therefore, it is crucial to develop a simple technique to characterise Dirac cones from optical absorption experiments alone.

In this work we present a theoretical description of the optical properties of generic 2D tilted Dirac cones by discussing the interband absorption of linearly polarised light (Section \ref{sec:Model}). We analyse the interband absorption (Section \ref{sec:Analysis}) discussing the key effects due to the tilt (or type) of Dirac cones (Section \ref{sec:Isotropic}) before including the effects of Fermi velocity anisotropy (Section \ref{sec:Anisotropic}). We discuss the wide range of absorption properties of Dirac cones with varying degrees of tilt (type-I, II and III) and anisotropy focusing on the polarisation-dependent behaviour at high frequencies (small Fermi energy) (Section \ref{sec:HighFreq}). We then explore the tunable frequency-dependent absorption at low frequency (large Fermi energy) (Section \ref{sec:LowFreq}). In these sections we show that the optical properties of a tilted Dirac cone can not be determined solely from its type. Finally, building on our analytical results we provide a systematic procedure to fully characterise tilted Dirac cones (i.e. their tilt and anisotropy parameters) from optical measurements alone (Section \ref{sec:Character}).

\begin{figure}
 \centering
 \includegraphics[width=0.48\textwidth]{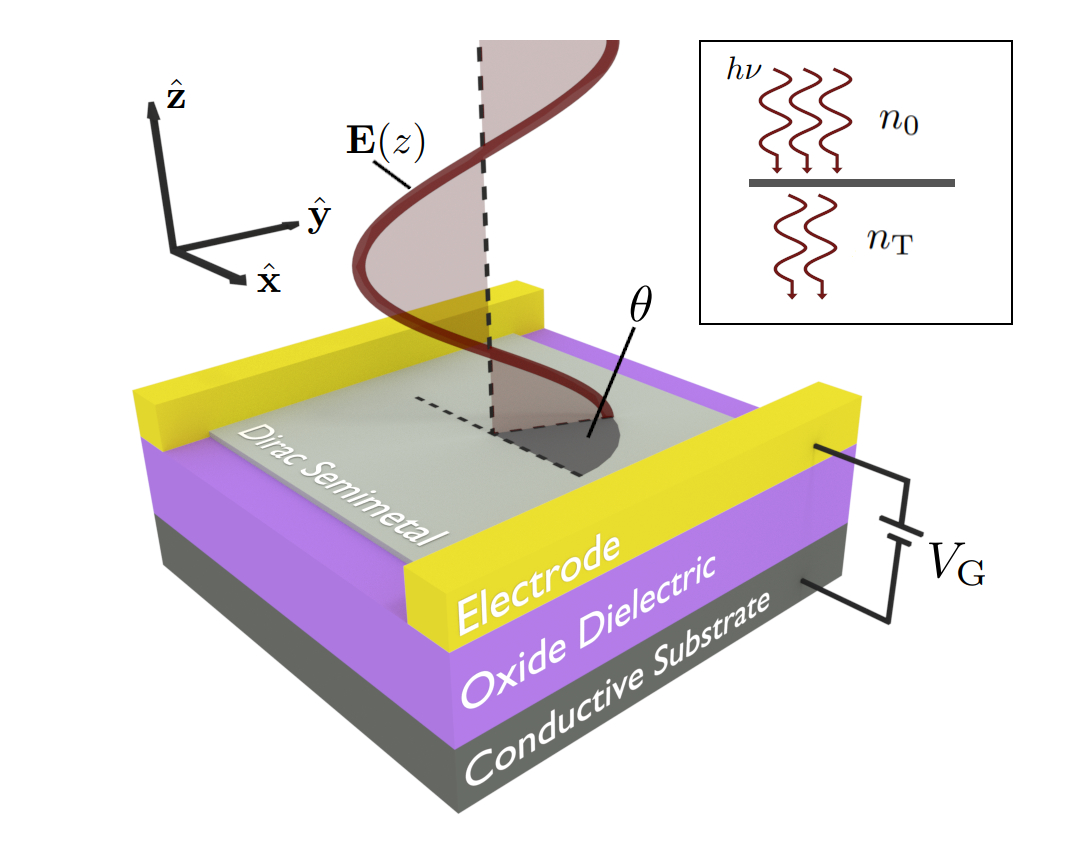}
 \caption{Schematic of suggested experimental setup for measuring the ratio of absorbed photons as $\mathcal{A} = n_\text{A}/n_0 = 1 - n_\text{T}/n_0$, where $n_0$, $n_\text{A}$ and $n_\text{T}$ are the incident, absorbed and transmitted photon densities per unit time respectively. The Fermi level $E_\text{F}$ of the Dirac semimetal can be changed via the back-gate voltage $V_\text{G}$. We consider monochromatic linearly polarised light with frequency $\nu$ incident normally to the sample with the electric field polarised at an angle $\theta$ from the $\hat{\textbf{x}}$ axis (corresponding to the tilt axis $q_x$ of the Dirac cone).}
 \label{fig:schem}
\end{figure}

\section{Model and absorption theory}
\label{sec:Model}

We consider a 2D DSM with a low energy band structure hosting two inequivalent tilted Dirac cones (valleys) where one of the cones is described by the effective Hamiltonian
\begin{equation}
\label{eq:DiracHamiltonian}
    H(\textbf{q}) = \hbar v_\text{F} \big( \gamma \eta q_x \sigma_0 + \eta q_x \sigma_x + q_y \sigma_y \big),
\end{equation}
where $\sigma_x$ and $\sigma_y$ are the Pauli matrices, $\sigma_0$ is the $2 \times 2$ identity matrix, $v_\text{F}$ is the Fermi velocity and $\textbf{q} = (q_x,q_y)$ is the wavevector deviation from the Dirac point. The Dirac cones are assumed to be tilted along the $q_x$ direction with a tilt parameter ($\gamma \geq 0$) and Fermi velocity anisotropy factor ($\eta > 0$). These parameters define three types of tilted Dirac cones: sub-critically tilted cones type-I, $\gamma < 1$ with closed elliptical isoenergy contours, critically tilted cones type-III, $\gamma = 1$ with open parabolic isoenergy contours and super-critically tilted cones type-II, $\gamma > 1$ with open hyperbolic isoenergy contours. Notice that the graphene limit is recovered in the case of no tilt ($\gamma = 0$) and isotropic Fermi velocity ($\eta = 1$). Time-reversal symmetry of the DSM directly yields the effective Hamiltonian of the second valley as $H^*(-\textbf{q})$. Diagonalizing Eq.\,(\ref{eq:DiracHamiltonian}) as $H\ket{\Psi_\pm} = E_\pm \ket{\Psi_\pm}$ gives eigenenergies
\begin{equation}
\label{eq:EigenEnergy}
    E_\pm(\textbf{q}) = \hbar v_\text{F} \widetilde{q} \big( \gamma \cos(\widetilde{\varphi}_\textbf{q}) \pm 1 \big),
\end{equation}
and eigenvectors
\begin{equation}
\label{eq:EigenVectors}
    \ket{\Psi_\pm(\textbf{q})} = \frac{1}{\sqrt{2}} \begin{pmatrix}
    \pm e^{-i\widetilde{\varphi}_\textbf{q}}\\
    1\end{pmatrix},
\end{equation}
for the valence ($-$) and conduction ($+$) bands. We use elliptical wavevector coordinates $\widetilde{q} = \sqrt{\eta^2 q_x^2 + q_y^2}$ and $\widetilde{\varphi}_\textbf{q} = \arctan(q_y/\eta q_x)$ allowing the Cartesian wavevectors to be written as $q_x = \widetilde{q} \cos(\widetilde{\varphi}_\textbf{q})/\eta$ and $q_y = \widetilde{q} \sin(\widetilde{\varphi}_\textbf{q})$, which reduce to standard polar representation ($q = \sqrt{q_x^2 + q_y^2}$ and $\varphi_\textbf{q} = \arctan(q_y/q_x)$) when $\eta = 1$. The system is placed in a back-gate configuration as shown in Fig.\,\ref{fig:schem} where the Fermi energy $E_\text{F} = \hbar v_\text{F} q_\text{F}$ can be tuned with a gate voltage. The Fermi energy is defined as positive $E_\text{F}>0$, the results of this paper will be unchanged for a Fermi level above or below the level crossing point. 

We consider a DSM illuminated at normal incidence by linearly polarised light with photon energy $h \nu$ and polarisation $\hat{\textbf{e}}_\theta = \cos(\theta)\hat{\textbf{x}} + \sin(\theta)\hat{\textbf{y}}$ at an angle $\theta$ to the $\hat{\textbf{x}}$ axis which corresponds to the tilt ($q_x$) axis of the Dirac cone as seen in Fig.\,\ref{fig:schem}. The electric field represents a time-dependent perturbation to the otherwise time-independent DSM Hamiltonian inducing transitions between the states $\ket{\Psi_\pm(\textbf{q})}$. In this work we ignore transitions within a single band (intraband) and focus on transitions between the valence and conduction bands (interband). The effects of intraband absorption have been considered previously for type-I Dirac cones\,\cite{Verma,Herrera}. In regard to type-II Dirac cones, as has been discussed in the context of super-critically tilted 3D Weyl cones, the intraband absorption is material-dependent and its analysis requires a detailed understanding of the Fermi surface\,\cite{Carbotte,Zyuzin} and the scattering mechanisms. 

The transition rate between the valence and conduction band at wavevector $\textbf{q}$ is found using Fermi's golden rule\,\cite{Anselm,HartmannPortnoi}
\begin{equation}
\label{eq:TransitionsRate}
    W_{\theta}(\nu,\textbf{q}) = \frac{2 \pi e^2 n_0}{c \nu} \mid \! v_{\text{cv},\theta}(\textbf{q}) \!\mid^2   \delta \big( \Delta E(\textbf{q}) - h \nu \big),
\end{equation}
where CGS units have been used, $\delta$ is the Dirac delta function, $\Delta E(\textbf{q}) = E_+(\textbf{q}) - E_-(\textbf{q})$ is the difference in energy between the valence and conduction bands at wavevector $\textbf{q}$ and $n_0$ is the incident photon density per unit time. The velocity matrix element (VME) 
\begin{equation}
\label{eq:VMEEq}
v_{\text{cv},\theta}(\textbf{q}) = \bra{\Psi_\pm(\textbf{q})} \hat{\textbf{e}}_\theta \cdot \textbf{v} \ket{\Psi_\mp(\textbf{q})}
\end{equation}
is given as the expectation value of the projection of the velocity operator $\textbf{v}$ along the polarisation vector $\hat{\textbf{e}}_\theta$ between the initial and final states belonging to the valence and conduction bands respectively. The velocity operator is given by
\begin{equation}
\label{eq:velOp}
    \textbf{v} = \frac{1}{\hbar} \boldsymbol{\nabla}_\textbf{q}H(\textbf{q})
\end{equation}
in the gradient approximation which is known to work in graphene near the Dirac cones apex\,\cite{MishaHartmann,HartmannPortnoi,Kresin2016}. In this work we consider weak excitation neglecting non-linear effects.

{\it (a) Vertical transitions.} Light induces vertical transitions between two states with the same $\textbf{q}$ and energy separation matching the photon energy, as accounted for by the Dirac delta function in Eq.\,(\ref{eq:TransitionsRate}). This condition determines the set of wavevectors $\textbf{q}$ involved in the photon absorption. The difference in energy between the valence and conduction band can be simply found from Eq.\,(\ref{eq:EigenEnergy})
\begin{equation}
\label{eq:contour}
    \Delta E(\textbf{q}) = 2 \hbar v_\text{F} \widetilde{q}.
\end{equation}
Interestingly, this band separation is independent of the tilt parameter ($\gamma$) meaning that the contour of allowed transitions $\widetilde{q} = h \nu/2\hbar v_\text{F}$ is unchanged for Dirac cones with varying $\gamma$ (from type-I to II to III).

\begin{figure*}
    \centering
    \includegraphics[width=1\textwidth]{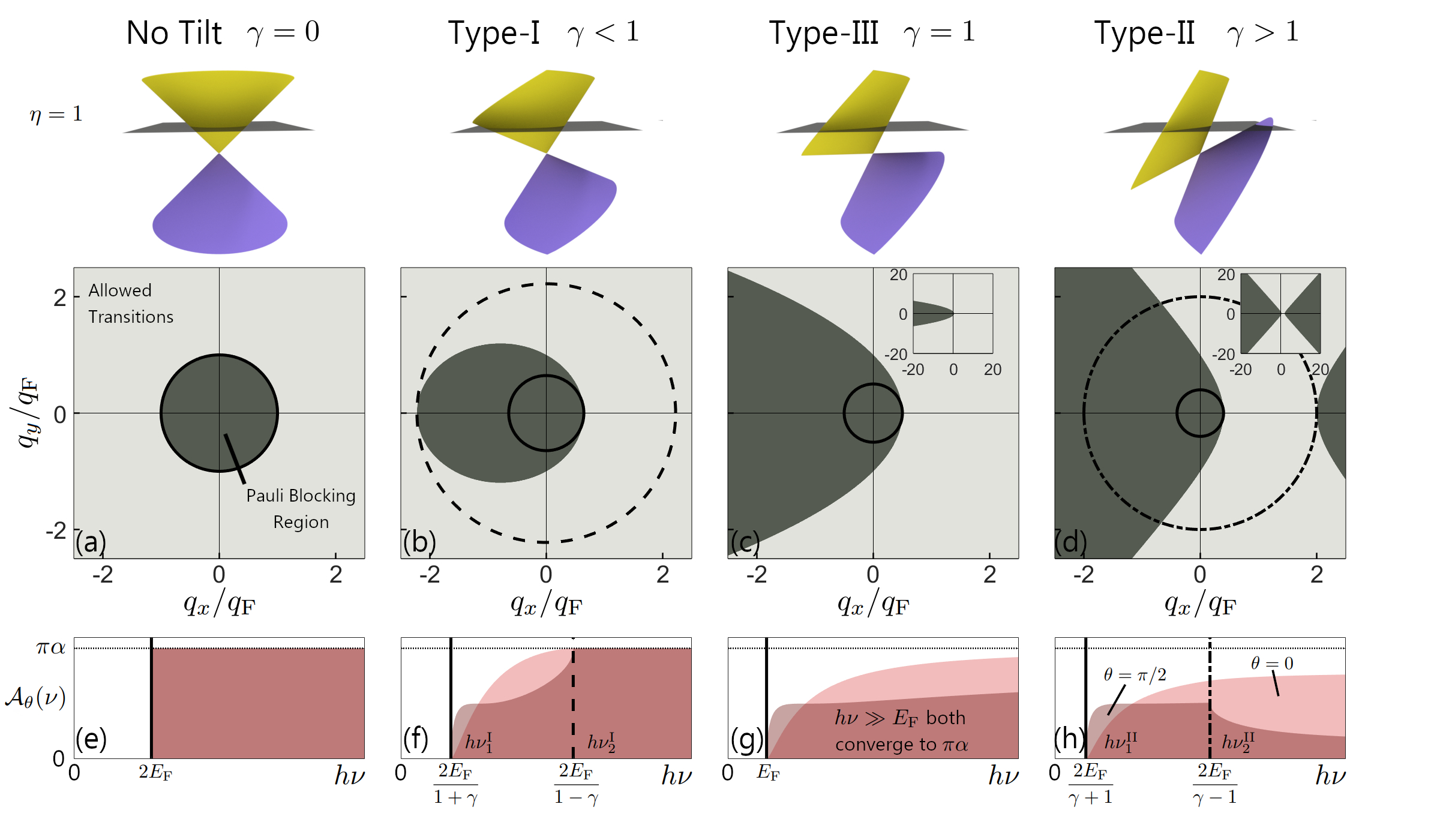}
    \caption{(a)-(d) Regions of allowed (light) and Pauli-blocked (dark) transitions in wavevector space for different values of $\gamma$ corresponding to non-tilted ($\gamma = 0$), type-I ($\gamma = 0.55$), type-III ($\gamma = 1.00$) and type-II ($\gamma = 1.50$) Dirac cones respectively. The Fermi level is scaled to the Fermi wavevector as $E_\text{F} = \hbar v_\text{F} q_\text{F}$ and the Fermi velocity is isotropic around the Dirac cones ($\eta = 1$). (e)-(h) The absorption $\mathcal{A}_\theta(\nu)$ is defined as the ratio of absorbed to incident photons with frequency $\nu$ polarised along $\theta = 0$ (light red) or $\theta = \pi/2$ (dark red) via interband processes and is plotted for zero temperature.}
    \label{fig:Absorption}
\end{figure*}




{\it (b) Momentum alignment.} Optical momentum alignment phenomenon is the anisotropic momentum distribution of photo-excited carriers with respect to the polarisation plane of the exciting linearly polarised light. It exists for interband transitions in various conventional semiconductors\,\cite{1982SvPhU..25..143Z} their quantum wells\,\cite{MomAlign2,MERKULOV1991371} and graphene\,\cite{MishaHartmann,PhysRevB.103.165411}. Here we show that momentum alignment is an ubiquitous phenomenon in tilted Dirac semimetals and analyse its dependence on their tilt and anisotropy parameters. 

For interband transitions in tilted Dirac cones Eq.\,(\ref{eq:EigenVectors}), (\ref{eq:VMEEq}) and (\ref{eq:velOp}) result in the following expression for the squared modulus of the velocity matrix element 
\begin{equation}
\label{eq:SelectionRules}
    \mid \! v_{\text{cv},\theta}(\textbf{q}) \! \mid^2 \: = v^2_\text{F} \big( \eta \cos(\theta) \sin(\widetilde{\varphi}_\textbf{q}) - \sin(\theta) \cos(\widetilde{\varphi}_\textbf{q}) \big)^2.
\end{equation}
This results in selection rules that block interband transitions for the states with wavevector parallel to the polarisation plane. Note that Eq.\,(\ref{eq:SelectionRules}) is written in terms of the elliptical wavevector angle $\widetilde{\varphi}_\textbf{q}$ instead of the true $\varphi_\textbf{q}$ angle, for brevity. Nevertheless, the selection rules still demonstrate momentum alignment. This is easier to see in the limit of isotropic Fermi velocity ($\eta = 1$) where the selection rules simplify to $\mid \!\! v_{\text{cv},\theta}(\textbf{q}) \!\! \mid^2 \: = v^2_\text{F} \sin^2(\varphi_\textbf{q} - \theta)$. The selection rules in Eq.\,(\ref{eq:SelectionRules}) are once again independent of the tilt parameter ($\gamma$) which demonstrates that momentum alignment seen in graphene remains a dominant effect in all tilted Dirac cones. Consequently, the results of Eqs.\,(\ref{eq:contour}) and (\ref{eq:SelectionRules}) show that the transition rate $W_{\theta}(\nu,\textbf{q})$ at wavevector $\textbf{q}$ given in Eq.\,(\ref{eq:TransitionsRate}) is also independent of the tilt parameter ($\gamma$) (from type-I to II to III) and depends only on the Fermi velocity anisotropy factor ($\eta$). 

{\it (c) Pauli blocking.} An electron can only be excited from the valence band to the conduction band if two criteria are met: the state in the valence band initially contains an electron and the state in the conduction band is initially empty. If either of these criteria are not met the transition is Pauli-blocked. This is accounted for by the Fermi-Dirac distributions for electrons ($e$) and holes ($h$), $f_e(E) = 1 - f_h(E) = \big[ 1 + \exp\big((E - \mu)/k_\text{B}T)\big]^{-1}$ with chemical potential $\mu$, temperature $T$ and Boltzmann constant $k_\text{B}$. We consider the limit of zero temperature (which is justified as long as $h \nu \gg k_\text{B} T$) where $\mu = E_\text{F}$ and the Fermi-Dirac distribution becomes a Heaviside step function $f_e(E) = \Theta(E_\text{F} - E)$. Finite temperatures would simply smear the Pauli blocked regions and therefore the frequency regimes of the absorption. In stark contrast to the contour of transitions and momentum alignment (in paragraphs {\it(a)} and {\it(b)} respectively) the regions of Pauli-blocked transitions depend on the tilt parameter ($\gamma$) as shown below.

{\it (d) Total absorption.} To obtain the total absorption (ratio of absorbed vs incident photons) of light we weight each transition by its rate ($W$) and Pauli blocking factors to calculate the density of absorbed $n_\text{A}$ photons per unit time. We then divide through by the density of incident photons per unit time $n_0$ to obtain the absorption
\begin{equation}
\label{eq:Absorb}
\begin{split}
    \mathcal{A}_{\theta}(\nu) = \frac{g_\text{s} g_\text{v}}{n_0 (2 \pi)^2}
    &\iint W_{\theta}(\nu,\textbf{q}) \times \\ &f_e \big(E_-(\textbf{q}) \big)f_h \big(E_+(\textbf{q}) \big) \text{d} \textbf{q},
\end{split}
\end{equation}
 where the area element in elliptical wavevector coordinates is $\text{d} \textbf{q} = \widetilde{q} \text{d} \widetilde{q} \text{d} \widetilde{\varphi}_\textbf{q}/\eta$. Thus far we have considered the absorption due to a single valley. Exploiting time-reversal symmetry, in the second valley the selection rules (square modulus of the VME) and the regions of Pauli blocked transitions can be found from the first valley by inverting the wavevector $\textbf{q} \to -\textbf{q}$. As a consequence, as the integration in Eq.\,(\ref{eq:Absorb}) sums over all wavevectors, both valleys yield the same absorption which can be accounted for by a valley degeneracy factor. Therefore, the factors $g_\text{v} = g_\text{s} = 2$ account for the valley and spin degeneracy respectively. The absorption will be discussed in terms of tilt ($\gamma$) and Fermi velocity anisotropy ($\eta$) for an incident photon polarisation ($\theta$). When discussing the absorption of specific Dirac cone types (I, II or III) this notation will be placed in the superscript of the absorption $\mathcal{A}_\theta$. 

\section{Analysis of absorption in tilted Dirac cones}
\label{sec:Analysis}

In this section, we analyse the absorption of light in materials hosting tilted Dirac cones. We calculate the absorption spectra for tilted Dirac cones without anisotropy in the Fermi velocity ($\eta = 1$) in Section \ref{sec:Isotropic} before generalising our model to include this anisotropy ($\eta \neq 1$) in Section \ref{sec:Anisotropic}. We then analyse the absorption spectra in the high frequency regime ($h \nu \gg  E_\text{F} $) in Section \ref{sec:HighFreq} and for photon energies of the order of the Fermi level ($h \nu \sim E_\text{F} $) in Section \ref{sec:LowFreq}.

\subsection{Absorption in isotropic Dirac cones}
\label{sec:Isotropic}

\subsubsection{Isotropic non-tilted Dirac cones}
We begin by briefly reviewing the well-known absorption of graphene that hosts Dirac cones without tilt ($\gamma = 0$) and with isotropic Fermi velocity ($\eta = 1$). As shown in Fig.\,\ref{fig:Absorption}(a) graphene has closed, circular isoenergy contours and Pauli blocking regions. For an incident photon of energy $h \nu$ wavevectors located on the contour $q = h \nu/2\hbar v_\text{F}$ contribute to absorption. In the low frequency regime ($h \nu \leq 2 E_\text{F}$, inside the solid contour) all possible transitions are Pauli-blocked leading to vanishing absorption. In contrast, in the high frequency regime ($h \nu > 2E_\text{F}$, outside the solid contour) all viable states are allowed to contribute to the absorption. In this high frequency regime we compute the integral in Eq.\,(\ref{eq:Absorb}) yielding the absorption $\mathcal{A}(\nu) = \pi \alpha \approx 2.3\%$ (where $\alpha \approx 1/137$ is the fine structure constant) for all $\theta$ as plotted in Fig.\,\ref{fig:Absorption}(e). This is the universal sheet absorbance of graphene\,\cite{GrapheneAbsTheory1,GrapheneAbsTheory2,GrapheneAbsExperiment} above the cut-off frequency $\nu = 2 E_\text{F}/h$.

\subsubsection{Isotropic type-I Dirac cones}
The absorption of type-I Dirac cones has been analysed in the context of two specific type-I DSMs: 8-$Pmmn$ Borophene\,\cite{Verma} and $\alpha$-(BEDT-TTF)$_2$I$_3$\,\cite{Nishine,Suzumura}. As demonstrated in Fig.\,\ref{fig:Absorption}(b) sub-critically tilted ($\gamma < 1$) isotropic ($\eta = 1$) type-I Dirac cones have closed elliptical isoenergy contours and Pauli-blocked regions. Like in graphene there exists a low frequency regime ($h \nu \leq h \nu_1^\text{I}$, inside the solid line) with $h \nu_1^\text{I} = 2 E_\text{F} /(1 + \gamma)$, where all transitions are Pauli-blocked and a high frequency regime ($h \nu > h \nu_2^\text{I}$, outside the dashed line) with $h \nu_2^\text{I} = 2 E_\text{F} /(1 - \gamma)$, where all transitions are allowed. These frequency regimes can be seen in the absorption spectra in Fig.\,\ref{fig:Absorption}(f) which clearly reduces to the graphene case as $\gamma = 0$.

The unique feature of type-I Dirac cones is the appearance of an intermediate frequency regime ($2 E_\text{F} /(1 + \gamma) < h \nu \leq 2 E_\text{F} /(1 - \gamma)$, outside the solid line and inside the dashed line in Fig.\,\ref{fig:Absorption}(b)) in which some states are Pauli-blocked. The absorption spectrum in this intermediate frequency regime, as seen in Fig.\,\ref{fig:Absorption}(f), can be explained by combining momentum alignment and Pauli blocking. First consider incident photons with energy just inside of the dashed regime; as can be seen in Fig.\,\ref{fig:Absorption}(b) most states with negative $q_x$ and small $q_y$ ($\varphi_\textbf{q} \approx \pi$) are Pauli-blocked. These are states that would most significantly contribute to absorption when the photons are polarised perpendicular to the tilt axis ($\theta = \pi/2$), reducing $\mathcal{A}^\text{I}_{\pi/2}$ compared to $\mathcal{A}^\text{I}_{0}$ as seen just to the left of the dashed line in Fig.\,\ref{fig:Absorption}(f). In contrast, for smaller energies just outside the solid line in Fig.\,\ref{fig:Absorption}(b) nearly all states are Pauli-blocked with the exception of states with positive $q_x$ and small $q_y$ ($\varphi_\textbf{q} \approx 0$), meaning $\mathcal{A}^\text{I}_{\pi/2}$ is large compared to $\mathcal{A}^\text{I}_{0}$, as seen just to the right of the solid line in Fig.\,\ref{fig:Absorption}(f). 

In summary, isotropic type-I Dirac cones have absorption spectra that only differs from the absorption of graphene in an intermediate frequency regime. In this regime the absorption is polarisation-dependent due to the elliptical Pauli blocking domain combined with momentum alignment effects. If the Fermi level is tuned to the level crossing point ($E_\text{F} = 0$) the absorption spectra of isotropic type-I cones will match that of graphene $\mathcal{A}^\text{I} = \pi \alpha$ without a low frequency cut-off. This is a consequence of the closed isoenergy contours in type-I Dirac cones and graphene alike, as no states are Pauli-blocked.

\subsubsection{Isotropic type-III Dirac cones}
Critically tilted ($\gamma = 1$) isotropic ($\eta = 1$) type-III Dirac cones have open parabolic isoenergy contours and Pauli-blocked regions, as seen in Fig.\,\ref{fig:Absorption}(c). Type-III Dirac cones can be understood as the limiting case of the type-I Dirac cone as $\gamma \to 1$. In this limit, the absorption in the low frequency regime ($h \nu \leq  E_\text{F}$, inside the solid line) behaves qualitatively the same as for type-I Dirac cones where all transitions are Pauli-blocked (see Fig.\,\ref{fig:Absorption}(g)). For all other frequencies ($h \nu > E_\text{F} $, outside the solid line) the absorption spectra is qualitatively the same as the intermediate frequency regime in type-I Dirac cones where the dashed line in Fig.\,\ref{fig:Absorption}(b) has been pushed to infinity. In the high frequency limit ($h \nu \gg E_\text{F}$) the parabolic Pauli-blocked domain blocks only a small portion of the wavevector space, as shown in the inset of Fig.\,\ref{fig:Absorption}(c). Consequently, in this regime the absorption for all polarisations converges to the polarisation-independent absorbance of type-I Dirac cones, so that $\mathcal{A}^\text{III} = \pi \alpha$.

\subsubsection{Isotropic type-II Dirac cones}
Super-critically tilted ($\gamma > 1$) isotropic ($\eta = 1$) type-II Dirac cones have open hyperbolic isoenergy contours and Pauli-blocked regions as can be seen in Fig.\,\ref{fig:Absorption}(d). Similarly to type-I Dirac cones there exists a low frequency regime ($h \nu \leq h \nu_1^\text{II}$, inside the solid line) with $h \nu_1^\text{II} = 2  E_\text{F} /(\gamma + 1)$, where all transitions are Pauli-blocked and an intermediate frequency regime ($h \nu_1^\text{II} < h \nu \leq h \nu_2^\text{II}$, outside the solid line and inside the dash-dot line) with $h \nu_2^\text{II} = 2 E_\text{F} /(\gamma - 1)$, where some states are Pauli-blocked by the first branch of the hyperbola. The absorption in these frequency regimes is qualitatively the same as for low and intermediate frequencies for type-I and type-III Dirac cones (see Fig.\,\ref{fig:Absorption}(h)).

Type-II Dirac cones have a unique high frequency regime different from all other tilted Dirac cones. As shown in Fig.\,\ref{fig:Absorption}d, in the high frequency regime ($h \nu > h \nu_2^\text{II}$, outside the dash-dotted line) the contour of contributing states is intersected by the second branch of the hyperbolic Pauli blocking region near the tilt axis ($\varphi_\textbf{q} \approx 0$). Due to the momentum alignment the second branch of Pauli blocking significantly reduces the absorption of high frequency photons polarised close to the perpendicular to the tilt axis ($\theta = \pi/2$). This is manifested as a kink just to the right of the dash-dotted line in Fig.\,\ref{fig:Absorption}(h). In the high frequency limit ($h \nu \gg  E_\text{F} $), as shown in the inset of Fig.\,\ref{fig:Absorption}(d), the hyperbolic Pauli blocking becomes frequency independent, blocking states around the $q_x$ axis and allowing transitions around the $q_y$ axis. In this limit the frequency independent absorption can be analytically expressed as
\begin{equation}
        \mathcal{A}^\text{II}_{0} = 2 \alpha \Bigg( \! \arcsin
        \bigg( \frac{1}{\gamma} \bigg) + \frac{1}{\gamma}\sqrt{1 - \frac{1}{\gamma^2}} \: \Bigg),
\end{equation}
and
\begin{equation}
        \mathcal{A}^\text{II}_{\pi/2} = 2 \alpha \Bigg( \! \arcsin
        \bigg(\frac{1}{\gamma}\bigg) - \frac{1}{\gamma}\sqrt{1 - \frac{1}{\gamma^2}} \: \Bigg),
\end{equation}
when $\eta = 1$. These high frequency polarisation-dependent limits are visible as saturation values in Fig.\,\ref{fig:Absorption}(h) when $h \nu \gg E_\text{F}$. 

To summarise, in the low and intermediate frequency regimes ($h \nu \leq h \nu_2^\text{II}$) type-II Dirac cones have qualitatively the same absorption as type-I and III Dirac cones. However, in stark contrast to all other cases, in the high frequency regime ($h \nu \gg E_\text{F}$) the absorption becomes increasingly polarisation-dependent as the tilt is increased. This high frequency, tilt-dependent, polarisation-sensitive absorption is a direct consequence of the open isoenergy contours and Pauli blocking regions only present in super-critically tilted type-II Dirac cones. 

\subsection{Absorption of anistropic tilted Dirac cones}
\label{sec:Anisotropic}
Up until this point we have discussed the absorption of tilted Dirac cones without discussing anisotropy in the Fermi velocity. In general, Dirac cones have a different Fermi velocity along the two axes $q_x$ and $q_y$ ($\eta \neq 1$).

As can be seen from Eqs.\,(\ref{eq:TransitionsRate}), (\ref{eq:SelectionRules}) and (\ref{eq:Absorb}), the absorption of photons polarised along $\theta = 0$ is $\eta^2$ times stronger than the absorption of photons polarised along $\theta = \pi/2$. Additionally, the change in the density of states at a fixed energy from the isotropic case can be described with the area element of the elliptical wavevectors as $\text{d} \textbf{q} = \widetilde{q} \text{d} \widetilde{q} \text{d} \widetilde{\varphi}_\textbf{q}/\eta$. As a consequence, the absorption of photons with an arbitrary polarisation of photon is multiplied by a factor $1/\eta$ compared to the isotropic case where $\text{d} \textbf{q} = q \text{d} q \text{d} \varphi_\textbf{q}$. Combining the selection rules and density of states in the anisotropic case, we can relate the absorption spectra of an anisotropic, tilted Dirac cone to its isotropic counterpart
\begin{equation}
\label{eq:anisotropicGeneral1}
    \mathcal{A}_{0}(\nu,\eta) = \eta \mathcal{A}_{0}(\nu,1),
\end{equation}
and 
\begin{equation}
\label{eq:anisotropicGeneral2}
\mathcal{A}_{\pi/2}(\nu,\eta) = \frac{1}{\eta} \mathcal{A}_{\pi/2}(\nu,1). 
\end{equation}
The case of general polarisation $\theta$ is derived in Appendix \ref{sec:AppendixAbsorption}. Therefore, the effect of anisotropy simply uniformly shifts the magnitude of absorption from the isotropic case depending on the polarisation.

\begin{figure}
 \centering
 \includegraphics[width=0.48\textwidth]{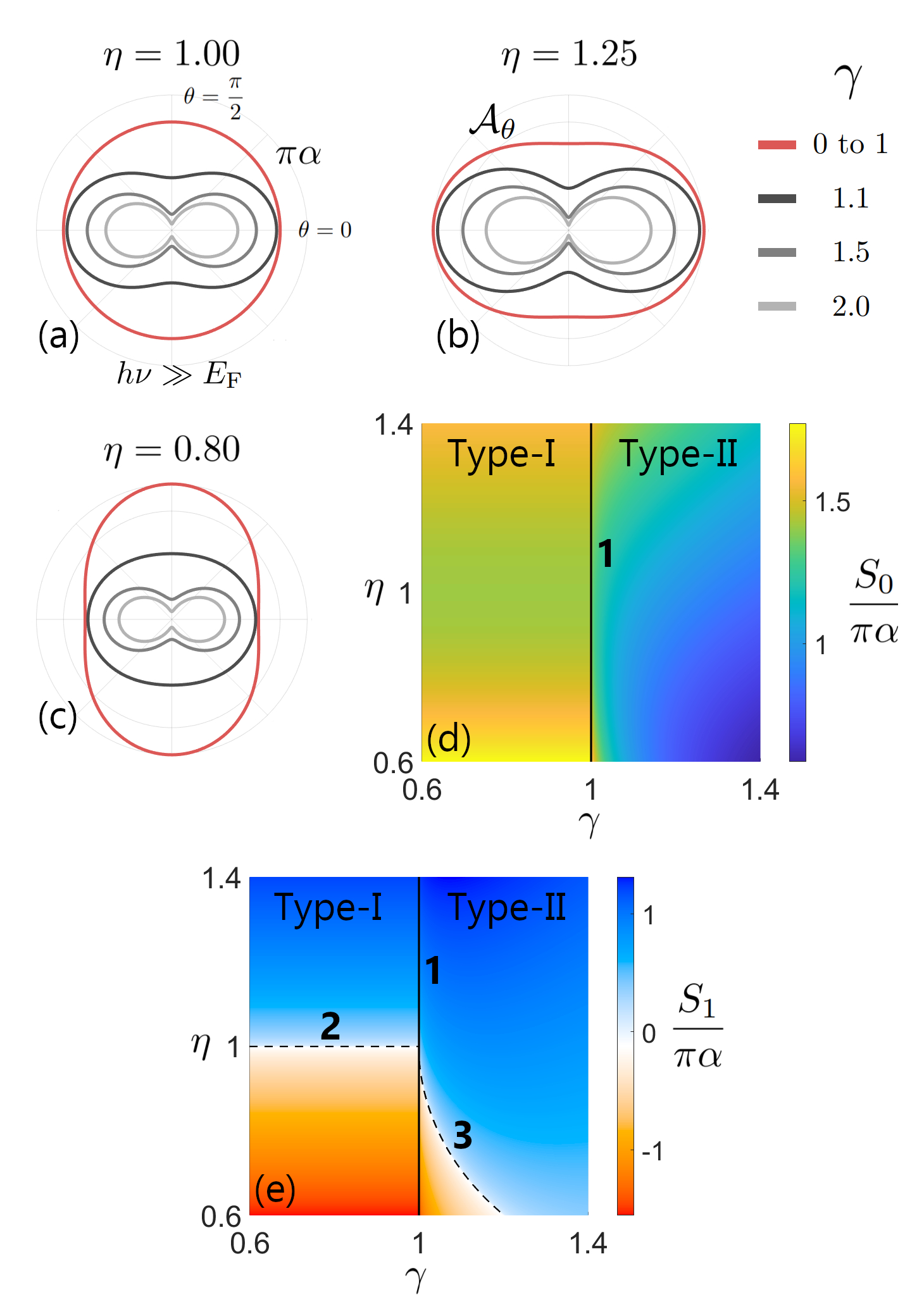}
 \caption{(a)-(c) The absorption $\mathcal{A}_\theta$  of light polarised along $\theta$ in the high frequency regime ($h \nu \gg E_\text{F}$) is plotted for three values of Fermi velocity anisotropy ($\eta$). The red lines correspond to type-I and III Dirac cones whilst the grey lines correspond to type-II Dirac cones. (d) and (e) The absorption Stokes parameters $S_0$ and $S_1$ as defined in Eqs.\,(\ref{eq:stokes1}) and (\ref{eq:stokes2}) are plotted for the parameter space $\gamma$ and $\eta$. In both figures contour 1 defines the boundary between type-I and type-II Dirac cones (corresponding to type-III Dirac cones). Contours 2 and 3 define the tilted Dirac cone parameters where in the high frequency regime, all polarisations of light are absorbed equally ($S_1 = 0$).}
 \label{fig:HighFreq}
\end{figure}

\subsection{Polarisation-dependent absorption of high frequency photons}
\label{sec:HighFreq}
We first focus our analysis on the absorption of photons in the high frequency regime ($h \nu \gg E_\text{F}$). In previous sections we demonstrated that in this frequency regime the absorption of linearly polarised light for all tilted Dirac cones tend to saturation values with varying dependence on polarisation ($\theta$), tilt parameter ($\gamma$) and Fermi velocity anisotropy ($\eta$). 

To explore the polarisation dependence, we focus on two linear polarisations of photons $\theta = 0$ (parallel to the corresponding tilt axis in momentum space $q_x$) and $\theta = \pi/2$ (perpendicular to the corresponding tilt axis) which will be shown to display the largest difference in absorption. We employ the parameters $\textbf{S} = \big( S_0, S_1 \big)$ defined as
\begin{equation}
\label{eq:stokes1}
        S_0 = \sqrt{\mathcal{A}^2_{0} + \mathcal{A}^2_{\pi/2}},
\end{equation}
and 
\begin{equation}
\label{eq:stokes2}
        S_1 = \text{sign}(\mathcal{A}_{0} - \mathcal{A}_{\pi/2} ) \sqrt{\big\lvert \mathcal{A}^2_{0} - \mathcal{A}^2_{\pi/2} \big\lvert},
\end{equation}
as the dimensionless absorption analogue of Stokes parameters. The first parameter $S_0$ measures the root mean square (RMS) of the absorbance for photons polarised along ($q_x$) and perpendicular ($q_y$) to the tilt axis axes and $S_1$ describes the asymmetry of absorption with a positive (negative) value indicating a greater absorbance of $q_x$ ($q_y$) polarised photons.

\subsubsection{Type-I and III Dirac cones} The high frequency absorbance of type-I and III Dirac cones depends on the Fermi velocity anisotropy ($\eta$) and polarisation ($\theta$) and is independent of tilt ($\gamma$)
\begin{equation}
\label{eq:TypeIAbs}
    \mathcal{A}^\text{I/III}_{\theta} = \frac{\pi \alpha}{2 \eta}\Big[ \big(1 + \eta^2 \big) - \big(1 - \eta^2 \big)\cos(2\theta)\Big],
\end{equation}
when $h \nu > 2 E_\text{F} /(1 - \gamma)$ (type-I) or $h \nu \gg  E_\text{F}$ (type-III). In Fig.\,\ref{fig:HighFreq}(a)-(c) (red lines) we plot the high frequency absorption for type-I and III Dirac cones for a variety of Fermi velocity anisotropy clearly showing that the absorption in this limit does not depend on the tilt (due to the closed regions of Pauli blocking). From Eq.\,(\ref{eq:TypeIAbs}) it can be seen that the largest difference of absorption between two perpendicular polarisations is when photons are polarised along ($\theta = 0$) and perpendicular ($\theta = \pi/2$) to the tilt axis.

\begin{figure*}
    \centering
    \includegraphics[width=1\textwidth]{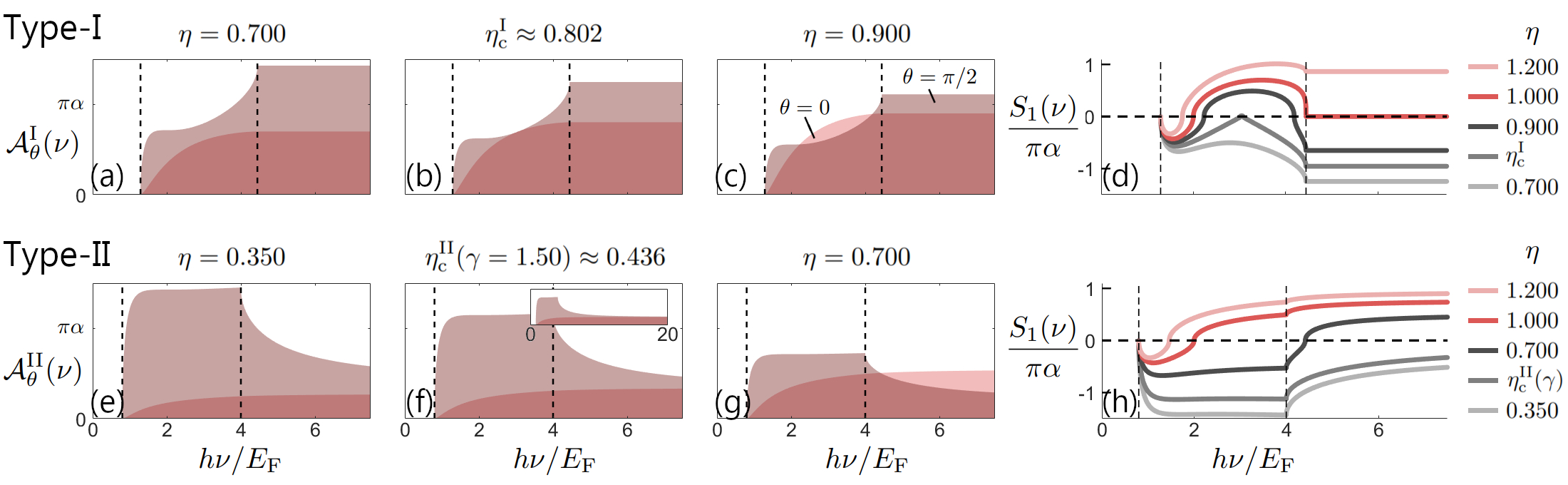}
    \caption{(a)-(c) and (e)-(g) The absorption spectra $\mathcal{A}_\theta(\nu)$ for light polarised along $\theta$ is plotted as a function of photon frequency for type-I ($\gamma = 0.55$) and type-II ($\gamma = 1.50$) Dirac cones, respectively, for several values of $\eta$. The inset of f shows the high frequency absorption tending to an isotropic ($S_1 = 0$) limit. The parameter $S_1$ as defined in Eq.\,(\ref{eq:stokes2}) defines the asymmetry of the absorption with respect to photon polarisation. It is plotted as a function of photon frequency for type-I and II Dirac cones in panels (d) and (h) respectively for a range of values $\eta$.}
    \label{fig:LowFreq}
\end{figure*}

 The RMS of the absorption is plotted in Fig.\,\ref{fig:HighFreq}(d) and is given by the expression $S_0 = \pi \alpha \sqrt{\eta^2 + (1/\eta)^2}$. The RMS absorption is invariant under the operation $\eta \to 1/\eta$ meaning there is no bias on increasing the Fermi velocity along or perpendicular to the tilt axis. The RMS absorption for type-I Dirac cones changes very slowly for a small amount of anisotropy. As seen in Fig.\,\ref{fig:HighFreq}(e) for type-I Dirac cones (top-left of contour ${1}$) $S_1$ is independent of tilt ($\gamma$) in the high frequency regime as previously seen in Eq.\,(\ref{eq:TypeIAbs}). Contour ${2}$ shows that when the Fermi velocity is isotropic ($\eta = 1$), photons of all polarisations are absorbed equally ($S_1 = 0$) matching the case of graphene.

\subsubsection{Type-II Dirac cones} 
\label{sec:HighFreqType2}
The high frequency absorption of type-II Dirac cones depends strongly on the tilt parameter ($\gamma$) due to the open, hyperbolic Pauli blocking regions. For a general photon polarisation ($\theta$), the absorption for type-II Dirac cones in the high frequency regime ($h \nu \gg E_\text{F}$) is given as
\begin{widetext}
\begin{equation}
\begin{split}
\label{eq:TypeIIAbs}
    \mathcal{A}^\text{II}_{\theta} = \frac{\alpha}{\eta} \Bigg\{ \Bigg[ \big( 1 &+ \eta^2\big)\arcsin \bigg( \frac{1}{\gamma} \bigg) - \frac{ 1-\eta^2}{\gamma}\sqrt{1 - \bigg( \frac{1}{\gamma}\bigg)^2} \Bigg] \:\:
    \\ - \Bigg[ \big( 1 &- \eta^2\big)\arcsin\bigg( \frac{1}{\gamma} \bigg) - \frac{ 1 + \eta^2}{\gamma}\sqrt{1 - \bigg( \frac{1}{\gamma}\bigg)^2} \Bigg]\! \cos(2\theta) \Bigg\}.
\end{split}
\end{equation}
\end{widetext}
In Fig.\,\ref{fig:HighFreq}(a)-(c) (all but red lines) the polarisation-dependent absorption plotted for a variety of tilt ($\gamma$) and Fermi velocity anisotropy ($\eta$). Like type-I and III Dirac cones the maximum and minimum absorption always occurs for photons polarised along ($\theta = 0$) or perpendicular to the tilt axis ($\theta = \pi/2$) corresponding to parallel and perpendicular to the tilt axis respectively.

As shown in Fig.\,\ref{fig:HighFreq}(d) (to the bottom-right of contour ${1}$), type-II Dirac cones have a significantly reduced RMS absorption $S_0$ compared to their type-I and III counterparts due to the hyperbolic Pauli blocking prohibiting many states from contributing to absorption. The parameter $S_1$ in Fig.\,\ref{fig:HighFreq}(e) shows the polarisation dependence of the absorption. This plot emphasises the two absorption mechanisms at play: the absorption anisotropy caused by hyperbolic Pauli blocking ($\gamma$ dependence) and from the anisotropic Fermi velocity ($\eta$ dependence). The parameters at which these two effects cancel (giving isotropic absorption $S_1 = 0$) are given by contour ${3}$ in Fig.\,\ref{fig:HighFreq}(e) which obeys
\begin{equation}
\label{eq:etaCrit2Ratio}
    \eta_\text{c}^\text{II}(\gamma) = \sqrt{\frac{C_-}{C_+}},
\end{equation}
where $C_\pm = (1/\pi)\big(\arcsin(1/\gamma) \pm (1/\gamma)\sqrt{1 - (1/\gamma)^2}\big)$. The majority blue colour in Fig.\,\ref{fig:HighFreq}(e) for type-II Dirac cones (bottom-right of contour ${1}$ and to the right contour ${3}$) demonstrates that in general, unless highly anisotropic, type-II Dirac cones will preferentially absorb photons polarised along the tilt axis. However, for parameters given by the contour in Eq.\,(\ref{eq:etaCrit2Ratio}) (on contour ${3}$) type-II Dirac cones can display no polarisation dependence; a property typically seen in graphene or isotropic type-I and III Dirac cones. Furthermore, in some type-II Dirac cones (left of contour ${3}$) photons polarised perpendicular to the tilt axis will be absorbed more favourably. This is a demonstration that simply classifying a Dirac cone as type-II is not enough information to understand its optical properties. 

\begin{figure*}
    \centering
    \includegraphics[width=.75\textwidth]{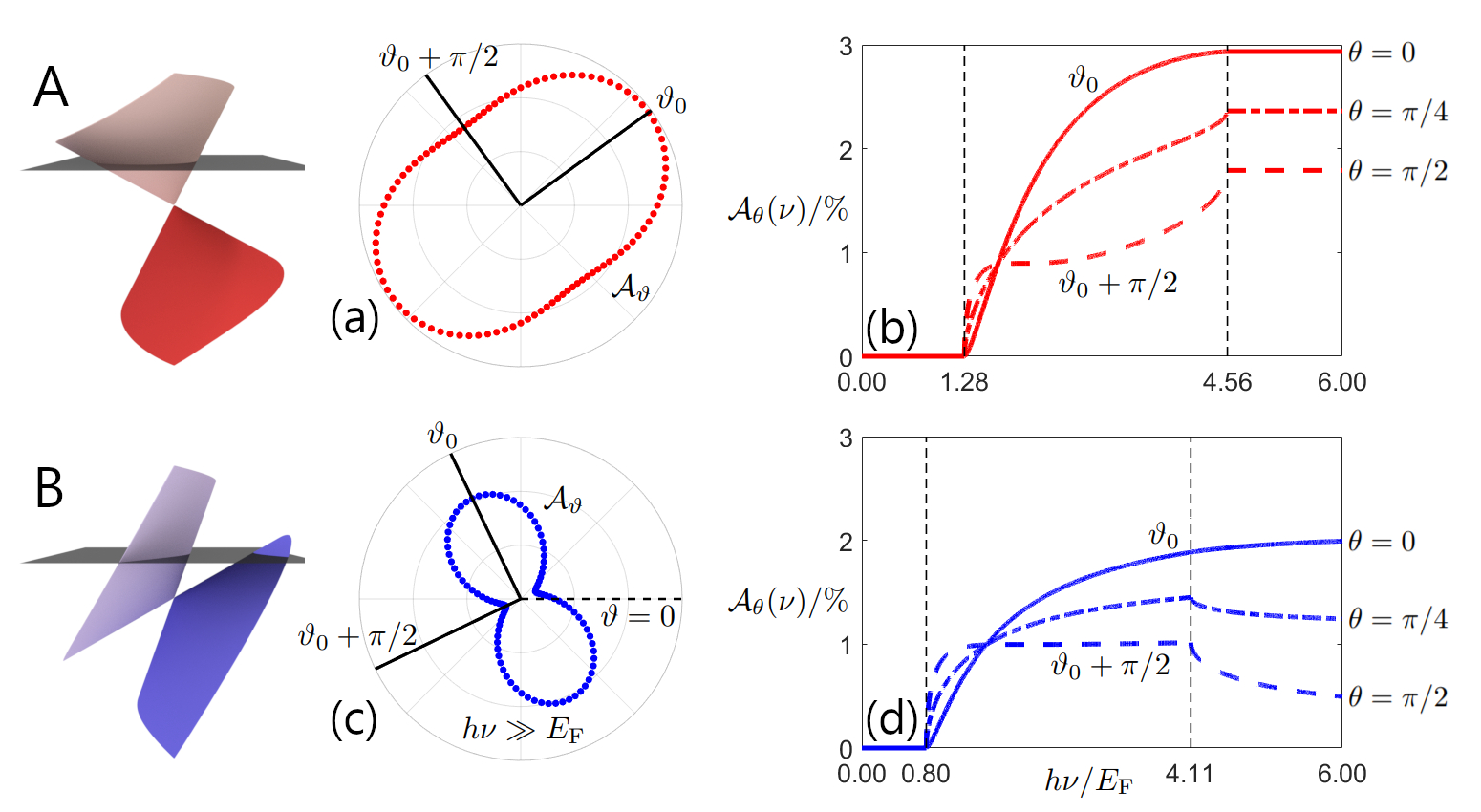}
    \caption{The absorption spectra is calculated in the high frequency regime ($h \nu \gg E_\text{F}$) for a polarisation angle $\vartheta$ for hypothetical Dirac cones A (a) and B (c). In both panels the polarisations corresponding to maximum or minimum absorption are denoted as $\vartheta_0$ and $\vartheta_0 + \pi/2$ respectively. For the two chosen polarisations $\vartheta_0$ or $\vartheta_0 + \pi/2$ the absorption spectra is calculated as a function of dimensionless photon energy $h \nu/E_\text{F}$ for Dirac cone A (b) and B (d). As justified in text these polarisations can be understood as the cases where the polarisation is along $\theta = 0$ or perpendicular $\theta = \pi/2$ to the tilt axis of the sample.}
    \label{fig:Character}
\end{figure*}

\subsection{Tunable polarisation-dependent low frequency response}
\label{sec:LowFreq}
In this section we focus on the qualitative differences in the absorption between different type-I and type-II Dirac cones for photon energies on the order of the Fermi level. It will be demonstrated that for both types of Dirac cone, there exist several qualitatively distinct absorption spectra, demonstrating that the Fermi velocity anisotropy is just as important as the tilt of the Dirac cone in determining the optical properties. Specifically, one of the few predicted 2D type-I DSMs, 8-$Pmmn$ Borophene, will be shown to straddle the border two different optical responses of type-I Dirac cones, possessing a unique absorption spectra. 

\subsubsection{Type-I Dirac cones with anisotropic Fermi velocity}
\label{sec:AnisotropicTypeI}
The absorption spectra for several different type-I Dirac cones with varying Fermi velocity anisotropy ($\eta$) and fixed tilt ($\gamma < 1$) is plotted in Fig.\,\ref{fig:LowFreq}(a)-(c). The frequency and polarisation dependence of the absorption is efficiently demonstrated with the asymmetry parameter $S_1(\nu)$ plotted in Fig.\,\ref{fig:LowFreq}(d). There are several distinct absorption responses in type-I Dirac cones. Firstly, when $\eta \lesssim 0.802$ (see Fig.\,\ref{fig:LowFreq}(a)) photons polarised perpendicular to the tilt axis ($\theta = \pi/2$) are for all frequencies absorbed more than photons of any other polarisations ($S_1 < 0$). Above this critical value, $\eta \gtrsim 0.802$ (see Fig.\,\ref{fig:LowFreq}(c)) the polarisation asymmetry $S_1$ changes sign twice. At the critical value $\eta = \eta^\text{I}_\text{c} \approx 0.802$, (see Fig.\,\ref{fig:LowFreq}(b)) $S_1$ is equal to zero for a single frequency. This is the largest value of $\eta$ for which light polarised perpendicular to the tilt axis is for all frequencies is absorbed more than all other polarisations. Equating the absorption of $\theta = 0$ and $\theta = \pi/2$ polarised light using expressions given explicitly in Appendix \ref{sec:AppendixAbsorption} yields the transcendental equation
\begin{equation}
\label{eq:transcendental}
        \arccos(\psi) = \frac{\eta^2 + 1}{\eta^2 - 1} \psi \sqrt{1 - \psi^2},
\end{equation}
where, $\psi = (1/\gamma)(2E_\text{F}/h \nu - 1)$. Equation (\ref{eq:transcendental}) has a single non-trivial solution when $\eta = \eta^\text{I}_\text{c} = \cot(x_0/2) \approx 0.802$, where $x_0$ is the first negative root of $x_0 = \tan(x_0)$. 

Surprisingly, the best-known 2D tilted type-I Dirac semimetal, 8-$Pmmn$ Borophene hosts Dirac cones with this exact value of $\eta = 0.802$\,\cite{Borophene2}. The absorption spectra of this material has been recently calculated\,\cite{Verma} with a tilt parameter of $\gamma = 0.46$ and appears qualitatively the same as Fig.\,\ref{fig:LowFreq}(b). At present it is not clear why 8-$Pmmn$ Borophene relaxes to this particular structure\,\cite{Borophene1}. This raises the question whether it would relax to a different structure if it were embedded in an anisotropic dielectric environment. If this was true it would pave the way to material engineering via electromagnetic environment.     

\subsubsection{Type-II Dirac cones with anisotropic Fermi velocity}
\label{sec:AnisotropicTypeII}
The absorption spectra for several different type-II Dirac cones with varying Fermi velocity anisotropy ($\eta$) and fixed tilt ($\gamma > 1$) is plotted in Fig.\,\ref{fig:LowFreq}(e)-(g) alongside the asymmetry parameter $S_1(\nu)$ in Fig.\,\ref{fig:LowFreq}(h). Like type-I Dirac cones there are two distinct absorption regimes. Firstly, when $\eta < \eta_\text{c}^\text{II}(\gamma)$ (see Fig.\,\ref{fig:LowFreq}(e)) photons polarised perpendicular to the tilt axis ($\theta = \pi/2$) are for all frequencies absorbed more strongly than photons of any other polarisations. In contrast, when $\eta > \eta_\text{c}^\text{II}(\gamma)$ (see Fig.\,\ref{fig:LowFreq}(g)) the polarisation asymmetry changes sign once with increasing frequency meaning that photons polarised along the tilt axis ($\theta = 0$) will be absorbed more than all other photons. Unlike type-I Dirac cones, in type-II Dirac cones the critical value of anisotropy $\eta_\text{c}^\text{II}(\gamma)$ (see Fig.\,\ref{fig:LowFreq}(f)) is a function of the tilt parameter. The value of $\eta_\text{c}^\text{II}(\gamma)$ was derived in Eq.\,(\ref{eq:etaCrit2Ratio}) and corresponds to contour ${3}$ in Fig.\,\ref{fig:HighFreq}(e). 

\section{Optical characterisation of tilted Dirac cones}
\label{sec:Character}
In this section we focus on optical characterisation of 2D materials hosting tilted Dirac cones. Our goal is to extract the material parameters $\gamma$ and $\eta$ from optical absorption experiments. Let us consider two hypothetical samples A and B corresponding to type-I and type-II Dirac cones as sketched in Fig.\,\ref{fig:Character}. The characterisation procedure requires measurements of the absorption spectra for polarisations aligned with the lattice axes of the sample $\theta = 0$ and $\pi/2$. Then these polarisation directions can be determined from optical experiments. The absorption measurements depend on the dimensionless photon energy $h \nu / E_\text{F}$. If only one lasing frequency $\nu$ is available in the experimental setup, the Fermi level can be tuned with respect to it by means of a back-gate voltage. Combining the absorption spectra and our analytical predictions we will then determine the anisotropy $\eta$ and tilt $\gamma$ parameters.  

\subsubsection{Determining the orientation of the sample}
To determine the orientation of the sample we propose two measurements. The first will reveal the direction of the $\theta =  0$ and $\pi/2$ polarisations, but will not distinguish one from the other. The second measurement will distinguish between these two polarisation directions. 

Firstly, the absorption scanning the polarisation directions in the high frequency regime $h \nu \gg E_\text{F}$ (see Fig.\,\ref{fig:Character}(a) and (c)). The experimental polarisation angle $\vartheta$ is measured from an arbitrary orientation of the sample and has no knowledge of the underlying lattice axes. As demonstrated in Section \ref{sec:HighFreq} the absorption maximum ($\vartheta_0$) and minimum ($\vartheta_0 + \pi/2$) correspond to polarisations aligned along or perpendicular to the tilt axis of the Dirac cone ($\theta = 0$ or $\pi/2$). Notice that there are two special cases where even tilted Dirac cones result in isotropic absorption at high frequency. These will be considered in Section \ref{sec:SpecialCases}.

To distinguish the $\theta = 0$ from the $\pi/2$ polarisation we propose a second measurement. For the two experimental polarisation angles $\vartheta_0$ and $\vartheta_0 + \pi/2$ the absorption should be measured as a function of the dimensionless photon energy $h \nu/E_\text{F}$ (see Fig.\,\ref{fig:Character}(b) and (d)). As seen in Section \ref{sec:Model}, the absorption spectra for $\theta = 0$ has no discontinuities in the gradient while the $\theta = \pi/2$ polarisation displays discontinuities at $h \nu_1/E_\text{F}$ and $h \nu_2/E_\text{F}$. From this point on-wards the orientation of the sample is known and the polarisation can be discussed only in terms of $\theta$.

\subsubsection{Distinguishing type-I from type-II Dirac cones}
\label{sec:CharacterA}
While scanning the absorption spectra with respect to the dimensionless photon energy for a generic polarisation angle $\theta \neq 0$, the absorption will show two discontinuities in the gradient at the critical frequencies $h \nu_1/E_\text{F}$ and $h \nu_2/E_\text{F}$. As derived in Appendix \ref{sec:AppendixAbsorption}, these critical frequencies do not depend on the value of $\theta$. Using our analytical expressions for the critical frequencies from Section \ref{sec:Model} we derive two tests. The first test will immediately allow us to distinguish type-I and type-II Dirac cones
\begin{equation}
\label{eq:TestII}
    \frac{E_\text{F}}{h \nu_1} + \frac{E_\text{F}}{h \nu_2} = \begin{cases}
    1 & \text{if type-I},\\
    \gamma > 1 & \text{if type-II}.
    \end{cases}
\end{equation} This test also provides the value of the tilt parameter $\gamma$ for type-II Dirac cones. If this test results in a type-I Dirac cone one can deduce $\gamma$ from 
\begin{equation}
\label{eq:TestI}
     \frac{E_\text{F}}{h \nu_1} - \frac{E_\text{F}}{h \nu_2} = \begin{cases}
    \gamma < 1 & \text{if type-I},\\
    1 & \text{if type-II}.
    \end{cases}
\end{equation} The first row of Eq.\,(\ref{eq:TestI}) was first derived in the context of 8-$Pmmn$ Borophene\,\cite{Verma}. In the specific case of non-tilted Dirac cones (e.g. graphene) one gets $\nu_1=\nu_2$ (as demonstrated in Fig.\,\ref{fig:Absorption}(e)) meaning Eq.\,(\ref{eq:TestI}) is equal to zero. Notably, type-III Dirac cones are described by the limit $E_\text{F}/h \nu_2 \to 0$ in Eqs.\,(\ref{eq:TestII}) and (\ref{eq:TestI}) and lie at the border between the type-I and II cases. 

Applying both of these tests to our hypothetical Dirac cones A and B in Fig.\,\ref{fig:Character}(b) and (d) yields the ratios $\gamma_\text{A} = 0.562$ for the type-I Dirac cone and $\gamma_\text{B} = 1.49$ for the type-II Dirac cone.

\subsubsection{Measuring the anisotropy parameters}
In order to determine the anisotropy parameter $\eta$ we return to the high frequency ($h \nu \gg E_\text{F}$) absorption measurements. An absorption measurement at two cross polarisations $\theta$ and $\theta + \pi/2$ (for example $\theta = 0$ and $\theta = \pi/2$) will lead to an average absorption $\overline{\mathcal{A}} = (\mathcal{A}_\theta + \mathcal{A}_{\theta + \pi/2})/2$ independent of $\theta$ as shown in Appendix \ref{sec:AppendixAbsorptionHighFreq}. Providing we have a type-I Dirac cone the anisotropy parameter $\eta$ can be extracted from the following expression
\begin{equation}
\label{eq:TypeIAniso}
     \eta^2 - \frac{2 \overline{\mathcal{A}}{}^{\text{I}}}{\pi \alpha}\eta + 1 = 0.
\end{equation}
In contrast, if the Dirac cone is type-II $\eta$ is obtained from
\begin{equation}
\label{eq:Type2FindEta}
    \eta^2 - \frac{\overline{\mathcal{A}}{}^{\text{II}}}{\pi \alpha C_+} \eta + \frac{C_-}{C_+} = 0,
\end{equation}
where $C_\pm$ are defined in Section \ref{sec:HighFreqType2} and crucially only depend on the tilt parameter $\gamma$ determined above in Eqs.\,(\ref{eq:TestII}) and (\ref{eq:TestI}). Both expressions yield two values of $\eta$. The correct value for $\eta$ corresponds to the larger root if $\mathcal{A}_0 > \mathcal{A}_{\pi/2}$ or the smaller root if $\mathcal{A}_0 < \mathcal{A}_{\pi/2}$.

In the case of the two hypothetical Dirac cones A and B, the average absorptions are $\overline{\mathcal{A}}{}^\text{A} = 2.36\%$ and $\overline{\mathcal{A}}{}^\text{B} = 1.19\%$. Applying the methods of this section yields the Dirac cone parameters $\eta_\text{A} = 1.28$ and $\eta_\text{B} = 1.15$.  

\subsubsection{Special cases: high frequency isotropic absorption}
\label{sec:SpecialCases}

As discussed in Section \ref{sec:HighFreq}, there are two special cases where the high frequency absorption ($h \nu \gg E_\text{F}$) does not depend on the polarisation of light even in the case of tilted, anisotropic Dirac cones. In these cases, the tilt parameter $\gamma$ should be found by following the method in Section \ref{sec:CharacterA} using the absorption spectra at an arbitrary polarisation direction. If the Dirac cone is type-I ($\gamma < 1$) then it follows that $\eta = 1$. If the Dirac cone is type-II ($\gamma > 1$) then it follows that $\eta = \eta^\text{II}_\text{c}(\gamma)$ which is found from Eq.\,(\ref{eq:etaCrit2Ratio}).

\section{Conclusion}
In this work we have given a comprehensive description of the absorption of linearly polarised light in 2D materials hosting tilted Dirac cones with tilt parameter $\gamma$. We show that super-critically tilted type-II Dirac cones ($\gamma > 1$) have high frequency absorption that depends on the tilt parameter $\gamma$. The reason for this is the open isoenergy contours in type-II Dirac cones yielding large regions of Pauli-blocked states causing certain polarisations to be absorbed stronger than others. This effect is in stark contrast to sub-critically tilted type-I ($\gamma < 1$) and critically tilted type-III ($\gamma = 1$) Dirac cones whose high frequency absorption shows no dependence on the tilt parameter. We also show that the absorption of tilted Dirac cones depends on the interplay of both the tilt parameter $\gamma$ and the Fermi velocity anisotropy $\eta$. For example, consider any type-I Dirac cone ($\gamma < 1$), if $\eta \lesssim 0.802$ it will for all frequencies absorb the polarisation of light aligned with the tilt axis of the cone stronger than all other polarisations. However, if $\eta \gtrsim 0.802$ the polarisation direction that is predominantly absorbed depends on frequency. Interestingly, the critical value of $\eta^\text{I}_\text{c} \approx 0.802$ corresponds to the best known type-I Dirac semimetal, $8-Pmmn$ Borophene. This observation raises the question of whether Borophene would relax to a different structure if it were embedded in an anisotropic dielectric environment. 

Using analytical results, we develop a recipe to characterise Dirac cones from optical measurements alone. In particular, we provide a systematic approach to determine the tilt parameter and Fermi velocity anisotropy. 

The exploration of 2D Dirac semimetals with tilted Dirac cones is an rapidly growing field with an ever rising number of suggested materials. One key feature demonstrated in our results is that for highly anisotropic structures ($\eta \sim 10$) the absorption ($\eta \pi \alpha \sim 20\%$) can be much higher than typically seen in conventional 2D semimetals with Dirac cones ($\pi \alpha \approx 2.3\%$). Dirac semimetals with tilted Dirac cones could thus be used as constituents for novel optoelectronic devices as they offer a tunable and highly polarisation-sensitive response. We hope that our work will guide the on-going search for thin-film materials with gate-tunable polarisation properties. 

\textbf{Note added in proof.} Some of the results of this paper have recently been obtained using the Kubo formalism in a preprint\,\cite{https://doi.org/10.48550/arxiv.2112.09392}, which appeared after the first arXiv version of our paper. The apparent similarity of the results shows the equivalence of the two approaches. 


\section*{ACKNOWLEDGMENTS}
This work was supported by the EU H2020-MSCA-RISE projects TERASSE (Project No. 823878) and DiSeTCom (Project No. 823728). A.W. is supported by a UK EPSRC PhD studentship (Ref. 2239575) and by the NATO Science for Peace and Security project NATO.SPS.MYP.G5860. E.M. acknowledges financial support from the Royal Society International Exchanges grant number IEC/R2/192166.

\appendix

\section{Analytic expressions for the absorption in arbitrarily tilted Dirac cones}
\label{sec:AppendixAbsorption}
In this section we derive the analytical absorption spectrum for type-I, II and III Dirac cones as a function of arbitrary photon frequency $\nu$ and polarisation $\theta$. Evaluating the integral in Eq.\,(\ref{eq:Absorb}) for type-I ($\gamma<1$) and type-III ($\gamma=1$) Dirac cones yields analytic expressions for the absorption. The absorption for a type-I DSM is
\begin{equation}
\begin{split}
\label{eq:TypeIa}
    \mathcal{A}^\text{I}_{\theta}(\nu) &= \Bigg( \frac{\mathcal{A}^\text{I}_{0}(\nu)  + \mathcal{A}^\text{I}_{\pi/2}(\nu)}{2} \Bigg) \\& \qquad + \Bigg( \frac{\mathcal{A}^\text{I}_{0}(\nu) - \mathcal{A}^\text{I}_{\pi/2}(\nu)}{2} \Bigg) \cos(2\theta),
\end{split}
\end{equation}
where
\begin{equation}
    \mathcal{A}^\text{I}_{0}(\nu) = \eta \pi \alpha \begin{cases}
    0, & h \nu \leq h \nu^\text{I}_1\\
    c^\text{I}_{0}(\nu), & h \nu^\text{I}_1 < h\nu \leq h \nu^\text{I}_2\\
    1, & h\nu > h \nu^\text{I}_2
    \end{cases},
\end{equation}
\begin{equation}
    \mathcal{A}^\text{I}_{\pi/2}(\nu) = \frac{\pi \alpha}{\eta} \begin{cases}
    0, & h \nu \leq h \nu^\text{I}_1\\
    c^\text{I}_{\pi/2}(\nu), & h \nu^\text{I}_1 < h\nu \leq h \nu^\text{I}_2\\
    1, & h\nu > h \nu^\text{I}_2
    \end{cases}.
\end{equation}
The boundaries between the frequency regimes are $h \nu^\text{I}_1 = 2 E_\text{F}/(1 + \gamma)$ and $h \nu^\text{I}_2 = 2 E_\text{F} /(1 - \gamma)$, while
\begin{equation}
    c^\text{I}_{0}(\nu) = \frac{1}{\pi} \Big( \arccos \psi_- - \psi_- \sqrt{1 - \psi_-^2 } \Big),
\end{equation}
\begin{equation}
\begin{split}
\label{eq:TypeIe}
    c^\text{I}_{\pi/2}(\nu) = \frac{1}{\pi}\Big( \arccos \psi_-  + \psi_- \sqrt{1 - \psi_-^2} \Big),
\end{split}
\end{equation}
with $\psi_- = (1/\gamma)(2E_\text{F}/h \nu -1)$ which is the same as $\psi$ in the main text. The type-III absorption spectrum can be calculated as the limiting case of the above Eqs.\,(\ref{eq:TypeIa}-\ref{eq:TypeIe}) for $\gamma \to 1$, where $h \nu^\text{I}_2 \to \infty$.

The absorption spectrum in type-II Dirac cones is
\begin{equation}
\begin{split}
\label{eq:TypeIIa}
    \mathcal{A}^\text{II}_{\theta}(\nu) &= \Bigg( \frac{\mathcal{A}^\text{II}_{0}(\nu) + \mathcal{A}^\text{II}_{\pi/2}(\nu)}{2} \Bigg) \\&\qquad + \Bigg( \frac{\mathcal{A}^\text{II}_{0}(\nu) - \mathcal{A}^\text{II}_{\pi/2}(\nu)}{2} \Bigg) \cos(2\theta),
\end{split}
\end{equation}
where
\begin{equation}
    \mathcal{A}^\text{II}_{0}(\nu) = \eta \pi \alpha \begin{cases}
    0, & h \nu \leq h \nu^\text{II}_1\\
    c^\text{IIa}_{0}(\nu), & h \nu^\text{II}_1 < h\nu \leq h \nu^\text{II}_2\\
    c^\text{IIb}_{0}(\nu), & h\nu > h \nu^\text{II}_2
    \end{cases},
\end{equation}
\begin{equation}
    \mathcal{A}^\text{II}_{\pi/2}(\nu) = \frac{\pi \alpha}{\eta} \begin{cases}
    0, & h \nu \leq h \nu^\text{II}_1\\
    c^\text{IIa}_{\pi/2}(\nu), & h \nu^\text{II}_1 < h\nu \leq h \nu^\text{II}_2\\
    c^\text{IIb}_{\pi/2}(\nu), & h\nu > h \nu^\text{II}_2
    \end{cases}.
\end{equation}
The boundaries between the frequency regimes are $h \nu^\text{II}_1 = 2 E_\text{F} /(\gamma + 1)$ and $h \nu^\text{II}_2 = 2 E_\text{F}/(\gamma - 1)$, while
\begin{equation}
    c^\text{IIa}_{0}(\nu) = \frac{1}{\pi}\Big( \arccos \psi_- - \psi_- \sqrt{1 - \psi_-^2} \Big),
\end{equation}
\begin{equation}
    c^\text{IIa}_{\pi/2}(\nu) = \frac{1}{\pi}\Big( \arccos \psi_- + \psi_- \sqrt{1 - \psi_-^2}\Big),
\end{equation}
\begin{equation}
\begin{split}
    c^\text{IIb}_{0}(\nu) = &\frac{1}{\pi}\Big( \arccos \psi_- - \arccos \psi_+ \\&- \psi_- \sqrt{1 - \psi_-^2} + \psi_+ \sqrt{1 - \psi_+^2}\Big),
\end{split}
\end{equation}
\begin{equation}
\label{eq:TypeIIf}
\begin{split}
    c^\text{IIb}_{\pi/2}(\nu) = &\frac{1}{\pi}\Big( \arccos \psi_- - \arccos \psi_+ \\&+ \psi_- \sqrt{1 - \psi_-^2} - \psi_+ \sqrt{1 - \psi_+^2} \Big),
\end{split}
\end{equation}
where $\psi_+ = (1/\gamma)(2E_\text{F}/h \nu +1)$.

\section{Polarisation independent optical properties}\label{sec:AppendixAbsorptionHighFreq} In this section of the appendix we discuss the average absorption for two cross polarisations for all tilted Dirac cones. As seen in Eqs.\,(\ref{eq:TypeIa}) and (\ref{eq:TypeIIa}) the average of the absorption for any two cross polarisations $\overline{\mathcal{A}} = (\mathcal{A}_\theta + \mathcal{A}_{\theta + \pi/2})/2$ is independent of $\theta$
\begin{equation}
    \overline{\mathcal{A}}(\nu) = \frac{\mathcal{A}_0(\nu) + \mathcal{A}_{\pi/2}(\nu)}{2}.
\end{equation}
In the high frequency regime ($h \nu \gg E_\text{F}$), the average absorption for type-I and type-II Dirac cones can be written as
\begin{equation}
\label{eq:AverageAbsType-I}
    \overline{\mathcal{A}}{}^\text{I} = \frac{\pi \alpha}{2 \eta}( 1 + \eta^2 ),
\end{equation}
\begin{equation}
\label{eq:AverageAbsType-II}
\begin{split}
    \overline{\mathcal{A}}{}^\text{II} &= \frac{\alpha}{\eta} \Bigg[ \big( 1 + \eta^2\big)\arcsin \bigg( \frac{1}{\gamma} \bigg) \\& \qquad\qquad\qquad - \frac{  1-\eta^2}{\gamma}\sqrt{1 - \bigg( \frac{1}{\gamma}\bigg)^2} \Bigg].
\end{split}
\end{equation}
It can be seen that in the case of a type-III Dirac cone ($\gamma = 1$), Eqs.\,(\ref{eq:AverageAbsType-I}) and (\ref{eq:AverageAbsType-II}) coincide.

\bibliography{References}
\end{document}